\documentclass[aps,prx,twocolumn,superscriptaddress,showpacs,floatfix]{revtex4-1}
\usepackage{amsmath,amssymb,graphicx}
\usepackage{wasysym}
\usepackage[utf8]{inputenc}
\usepackage[T1]{fontenc}
\usepackage{xcolor}
\allowdisplaybreaks
\usepackage{rotating} 

\usepackage{textcomp}

\usepackage{bm}

\IfFileExists{siunitx.sty}{\usepackage{booktabs,siunitx}}{}

\usepackage{color}
\definecolor{LinkColor}{rgb}{0.256,0.439,0.588}
\usepackage{hyperref}

\renewcommand{\vec}[1]{\mathbf{#1}}

\usepackage{pifont}

\newcommand{\La}{\line (1,0  ){12}}
\newcommand{\Lb}{\line (0,10 ){11}}

\newcommand{\Ld}{\line (-1,0){12}}
\newcommand{\Le}{\line (0,-10){12}}

\newcommand{\C} {\circle*{4}}

\newcommand{\LaT}{\rule[-1pt]{0.4cm}{0.2em}}  
\newcommand{\LdT}{\rule[-1pt]{0.4cm}{0.2em}}  
\newcommand{\LbT}{\rotatebox{90}{\rule[-1pt]{0.4cm}{0.2em}}}  
\newcommand{\LeT}{\rotatebox{90}{\rule[-1pt]{0.4cm}{0.2em}}}  

\newcommand{\pA}{\put(-3,-10)}
\newcommand{\pB}{\put(9,-10)}
\newcommand{\pC}{\put(9,0)}

\newcommand{\pZ}{\put(-3,0)}

\newcommand{\pAT}{\put(-4,-10)} 
\newcommand{\pBT}{\put(8.2,-10)}  

\newcommand{\rhomb}{
  \pA{\C}\pB{\C}\pZ{\C}\pC{\C}
 }

\newcommand{\rhombH}{
  \begin{picture}(22,10)(-8,-6)
    \pA{\LaT}\pB{\Lb}\pZ{\Le}\pZ{\LdT}
    \rhomb
  \end{picture}
}
\newcommand{\rhombV}{
  \begin{picture}(22,10)(-8,-6)
   \pA{\La}\pBT{\LbT}\pAT{\LeT}\pC{\Ld}
    \rhomb
  \end{picture}
}

\newcommand{\pBTs}{\put(5,-10)}  
\newcommand{\LbTs}{\rotatebox{90}{\rule[-1pt]{0.2cm}{0.05cm}}}  
\newcommand{\pATs}{\put(0.5,-10)} 
\newcommand{\LeTs}{\rotatebox{90}{\rule[-1pt]{0.2cm}{0.05cm}}}  
\newcommand{\pAs}{\put(0.5,-9)}
\newcommand{\LaTs}{\rule[-1pt]{0.2cm}{0.05cm}}  

\newcommand{\sqt}{
  \begin{picture}(7,4)(-0.5,-9)
  \pAs{\LaTs}\pBTs{\LbTs}\pATs{\LeTs}
  \end{picture}
}

\newcommand{\pZs}{\put(0.5,-4.5)}
\newcommand{\LdTs}{\rule[-1pt]{0.2cm}{0.05cm}}  

\newcommand{\sqb}{
  \begin{picture}(6,4)(-0.5,-9)
\pZs{\LdTs}\pBTs{\LbTs}\pATs{\LeTs}
  \end{picture}
}

\newcommand{\sql}{
  \begin{picture}(6,4)(-0.5,-9)
\pZs{\LdTs}\pATs{\LeTs}\pAs{\LaTs}
  \end{picture}
}

\newcommand{\sqr}{
  \begin{picture}(6,4)(-0.5,-9)
\pAs{\LaTs}\pBTs{\LbTs}\pZs{\LdTs}
  \end{picture}
}

\newcommand{\bea}{\begin{eqnarray}}
\newcommand{\eea}{\end{eqnarray}}

\newcommand{\vdimer}{{\vrule height0.2cm width0.05cm depth0pt}}
\newcommand{\hdimer}{{\hrule height0.05cm width0.2cm depth0pt}}

\newcommand{\mdimer}{\vbox{\hdimer \vskip 0.0625cm}}

\newcommand{\verdimers}{\hbox{\vdimer \hskip 0.1cm \vdimer}}
\newcommand{\hordimers}{\hbox{\vbox{\hdimer \vskip 0.1cm \hdimer}}}

\newcommand{\overdimer}{\hbox{\hskip 0.05cm \vdimer }}
\newcommand{\ohordimer}{\hbox{\vbox{\mdimer}}}

\begin{document}

\title{Fully packed quantum loop model on the square lattice: \\Phase diagram and application for Rydberg atoms}

\author{Xiaoxue Ran}
\affiliation{Department of Physics and HKU-UCAS Joint Institute of Theoretical and Computational Physics,The University of Hong Kong, Pokfulam Road, Hong Kong SAR, China}

\author{Zheng Yan}
\email{zhengyan@hku.hk}
\affiliation{Department of Physics and HKU-UCAS Joint Institute of Theoretical and Computational Physics,The University of Hong Kong, Pokfulam Road, Hong Kong SAR, China}
\affiliation{State Key Laboratory of Surface Physics, Fudan University, Shanghai 200433, China}

\author{Yan-Cheng Wang}
\affiliation{Beihang Hangzhou Innovation Institute Yuhang, Hangzhou 310023, China}

\author{Junchen Rong}
\affiliation{Institut des Hautes \'Etudes Scientifiques, 91440 Bures-sur-Yvette, France}

\author{Yang Qi}
\affiliation{State Key Laboratory of Surface Physics, Fudan University, Shanghai 200433, China}
\affiliation{Center for Field Theory
	and Particle Physics, Department of Physics, Fudan University,	Shanghai 200433, China}
\affiliation{Collaborative Innovation Center of Advanced Microstructures, Nanjing 210093, China}

\author{Zi Yang Meng}
\email{zymeng@hku.hk}
\affiliation{Department of Physics and HKU-UCAS Joint Institute of Theoretical and Computational Physics,The University of Hong Kong, Pokfulam Road, Hong Kong SAR, China}

\begin{abstract}
The quantum dimer and loop models attract great attentions, partially because the fundamental importance in the phases and phase transitions emerging in these prototypical constrained systems, and partially due to their intimate relevance toward the on-going experiments on Rydberg atom arrays in which the blockade mechanism naturally enforces the local constraint. Here we show, by means of the sweeping cluster quantum Monte Carlo method, the complete ground state phase diagram of the fully packed quantum loop model on the square lattice. We find between the lattice nematic (LN) phase with strong dimer attraction and the staggered phase (SP) with strong dimer repulsion, there emerges a resonating plaquette (RP) phase with off-diagonal translational symmetry breaking. Such a quantum phase is separated from the LN via a first order transition and from the SP by the famous Rokhsar-Kivelson point. Our renormalization group analysis reveals the different flow directions, fully consistent with the order parameter histogram in Monte Carlo simulations. The realization and implication of our phase diagram in Rydberg experiments are proposed.
\end{abstract}

\date{\today}
\maketitle

\section{Introduction}
Quantum dimer and loop models and their classical cousins are the prototypical constrained many-body systems~\cite{fisherStatistical1961,kasteleynThe1961,temperley1961dimer,fisherStatisticalII1963,kivelsonTopology1987,davidCoulomb2003,ardonneTopological2004,aletInteracting2005,aletClassical2006,moessnerResonating2001,moessnerPhase2001,moessnerIsing2001,fradkinBipartite2004,moessnerQuantum2011,dabholkarReentrance2022,charrierPhase2010,charrierGauge2008}. In 2D lattices, the quantum dimer model (QDM) usually refers to the local constraint with dimer covering of one dimer per site and the quantum loop model (QLM) two dimers per site. The QDM and QLM can be viewed as the incarnation of the resonating valence bond wave functions~\cite{kivelsonTopology1987,andersonResonating1987,rokhsarSuperconductivity1988} and the low-energy effective model of frustrated magnets~\cite{moessnerResonating2001,moessnerPhase2001,moessnerIsing2001}, and they offer the clear realization of the lattice gauge theory and conformal quantum criticalities~\cite{readLarge1991,wenMean1991,rodolfoSpontaneous1991,ivanovVortexlike2004,ralkoZero2005,ralkoDynamics2006,ralkoCrystallization2007,charrierGauge2008}. In recent years, these constrained quantum many-body models attract broad research interests since they can be realized in the Rydberg atom arrays trapped in optical tweezers~\cite{saffmanQuantum2010, browaeysMany2020,bernienProbing2017}, in which each dimer is identified with an atom excited to a Rydberg state by laser pumping~\cite{samajdarQuantum2021,verresenPrediction2021,Yue2021order,yanTriangular2022}. The observations of quantum phase transitions and the signature of topological orders from such experiments~\cite{schollQuantum2021,ebadiQuantum2021,satzingerRealizing2021,semeghiniProbing2021} have posted the questions of the complete and precise theoretical understanding of the phase diagrams of QDM and QLM to the community.

However, the question is by no means easy to answer. The precise computation of the physical properties of QDM and QLM are difficult besides few limits, such as the Rokhsar-Kivelson (RK) point~\cite{kivelsonTopology1987,henleyRelaxation1997,laeuchliDynamical2008} or exact diagonalization and DMRG for small clusters, and Green Function Monte Carlo simulations at intermediate system sizes~\cite{shannonCyclic2004,ivanovVortexlike2004,ralkoZero2005,ralkoDynamics2006,ralkoCrystallization2007,roychowdhuryZ22015,platMagnetization2015}. Recently, thanks to the invention of the sweeping cluster quantum Monte Carlo (QMC) method~\cite{yanSweeping2019,yanTopological2021,ZY2020improved,yanTriangular2022,yanEmergent2023}, which solves these constrained models in the path integral via efficient Monte Carlo update scheme respecting the local constraints, the ground state phase diagrams and the low-energy excitations therein of QDM on the square~\cite{yanWidely2021,yanHeight2022,dabholkarReentrance2022,yanTriangular2022}, triangular~\cite{yanTopological2021} and of QLM on triangular lattices~\cite{yanFully2022} have been obtained with controlled finite size analysis towards the thermodynamic limit (TDL). Meanwhile, the phase diagrams of the QDM and QLM on other 2D lattices, as simple as QLM on square lattice, are still largely unknown and their implications to the on going experiments in Rydberg arrays are yet to be explored.

\begin{figure*}[htp!]
	\centering
	\includegraphics[width=0.8\textwidth]{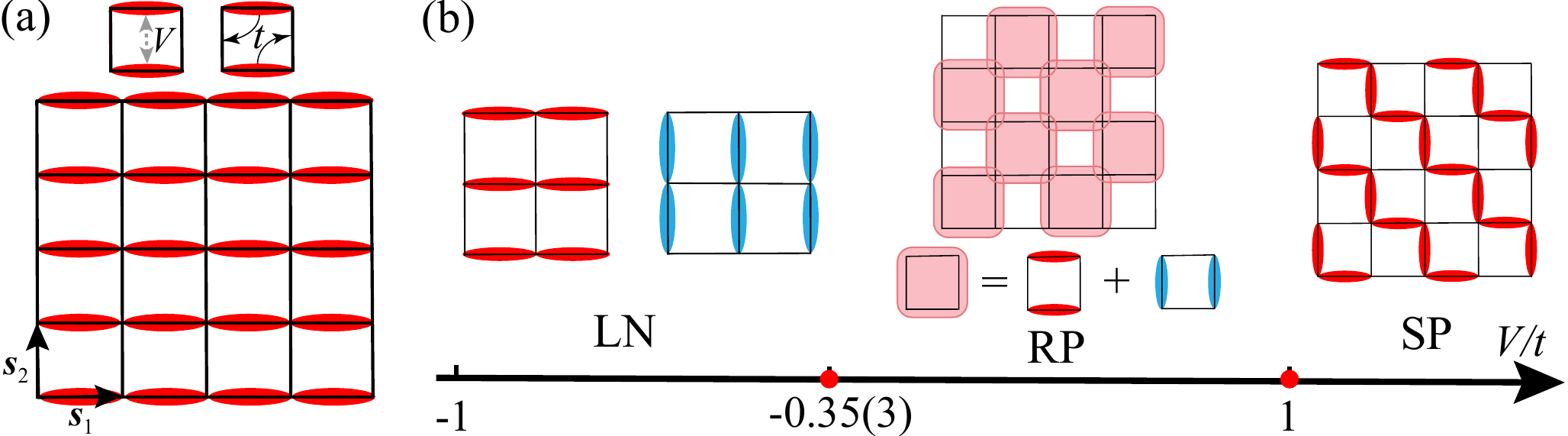}
	\caption{Fully packed quantum loop model on the square lattice. (a) Schematic representation of the QLM; $\mathbf{s}_{1}$ and $\mathbf{s}_{2}$ are the primitive vectors. The dimer configuration shown is one of the two LN patterns, with fully packed loops along the $\mathbf{s}_{1}$ direction. (b) Phase diagram of the QLM obtained from our simulations. The left subfigure illustrates the two LN dimer configurations. The first row of the middle subfigure is one of the two RP patterns and the second row is the resonating dimer pair within one plaquette. The first-order phase transition between the LN and RP states occurs at $V=-0.35(3)$. The right subfigure shows one representative SP state with $V>1$. The SP states have extensive ground state degeneracy $\sim 4\times 2^{p}$ ($p \propto L$ as discussed in Ref.~\cite{shannonCyclic2004}).}
	\label{fig:fig1}
\end{figure*}

Here, we answer the question of the phase diagram of the QLM on the 2D square lattice with the sweeping cluster QMC. Via the finite size scaling toward the TDL and the renormalization group analysis of the effective height field action~\cite{kivelsonTopology1987,ardonneTopological2004}, we find there exists a lattice nematic (LN) phase with strong dimer attraction and a staggered phase (SP) with strong dimer repulsion, and in between, there further emerges a resonating plaquette (RP) phase with off-diagonal long-range order that breaks the lattice translational symmetry. Such a quantum phase is separated from the LN via a first order transition and from the SP by the RK point. This intermediate RP phase, diagnosed not by the dimer correlation but by the off-diagonal $t$ term [see the Hamiltonian Eq.~\eqref{eq:eq1}] correlation, has both theoretical interests with its dangerously irrelevant $\cos(4\pi h)$ operator at the RK point similar to the second length scale in the deconfined quantum criticality~\cite{senthilQuantum2004,shaoQuantum2016}, and more importantly, the experimental detectable signature in the Rydberg atom arrays on a checkerboard lattice. Our results therefore respond to the urgent question from the fast progress in experiments and can be used to guide future ones.

\section{Model and methods}
The Hamiltonian of QLM on a square lattice is defined as
\begin{eqnarray}
  H=&-t&\sum_{plaq} \left(
  \left|\rhombV\right>\left<\rhombH\right| + h.c.
  \right) \nonumber \\
  &+V&\sum_{plaq}\left(
  \left|\rhombV\right>\left<\rhombV\right|+\left|\rhombH\right>\left<\rhombH\right|
  \right),
\label{eq:eq1}
\end{eqnarray}
where the summation covers all plaquettes and as shown in Fig.~\ref{fig:fig1}(a). The local constraint is implemented such that there must be two dimers touching every lattice site in every dimer configuration. The kinetic term---$t$ term---changes the dimer covering of flippable plaquettes, and the potential term---$V$ term---is repulsive ($V>0$) or attractive ($V<0$) between dimers facing each other on a plaquette, and we also allow the configurations with three dimers on a plaquette. The RK point is located at $V/t=1$ and described by an emergent $U(1)$ symmetry in the effective height model description~\cite{kivelsonTopology1987,ardonneTopological2004} (see below). We set $t=1$ as the unit of energy in our simulations.

\begin{figure}[tb]
	\centering
	\includegraphics[width=1\columnwidth]{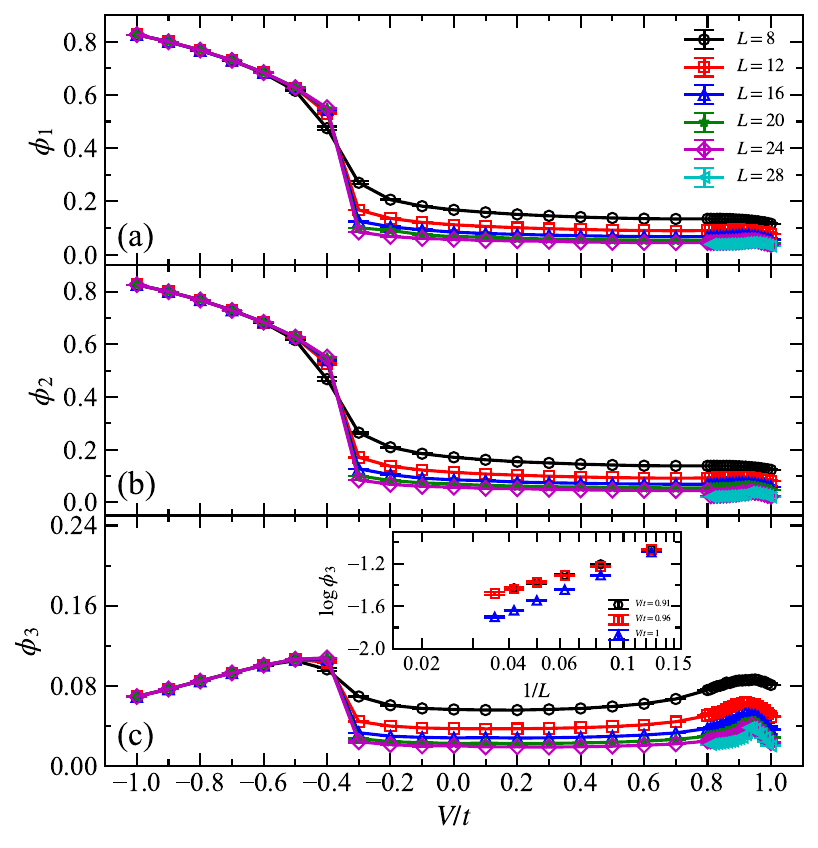}
	\caption{Quantum phase transitions in the square lattice QLM. (a), (b), and (c) are the dimer order parameters $\phi_{1}, \phi_{2}$, and $\phi_{3}$ as a function of $V$. The LN-RP first-order transition occurs at $V=-0.35(3)$. The inset of (c) shows the instability of the rotational symmetry breaking inside each plaquette close to the RK point, is a finite size effect, and in a log-log plot, $\phi_3$ at $V\sim 0.9$ goes to zero in a power-law with $L$.}
	\label{fig:fig2}
\end{figure}

The sweeping cluster QMC approach~\cite{yanSweeping2019,ZY2020improved,yanTopological2021,yanTriangular2022} employed in this work is a new method developed by us, which works well in constrained quantum lattice models~\cite{yanSweeping2019,yanTopological2021,ZY2020improved}.
Prior to sweeping cluster QMC, to solve the QDM or QLM types of constrained models, one had to rely on either exact diagonalization of small systems, or variational approaches such as DMRG that suffer from finite-size effects on the cylindrical geometry~\cite{roychowdhuryZ22015}, or the projector Monte Carlo approaches, which include the Green's function~\cite{ivanovVortexlike2004,ralkoZero2005,ralkoDynamics2006,vernayIdentification2006,ralkoCrystallization2007,platMagnetization2015} and diffusion Monte Carlo schemes~\cite{ofsContinuous2005,ofsPlaquette2006}, or sampling directly in height space and throwing away the unconstrained configurations~\cite{banerjeeThe2013,banerjeeInterfaces2014,banerjeeFinite2016}. These projector Monte Carlo methods obey the geometric constraints but are not efficient away from the RK point~\cite{ofsRandom2005} and only work at $T=0$. Furthermore, there does not exist any cluster update for the projector methods. On the contrary, the sweeping cluster algorithm is based on the world-line Monte Carlo scheme~\cite{ofsQuantum2002,aletGeneralized2005,aletInteracting2005} to sweep and update layer by layer along the imaginary time direction so that the local constraints (gauge field) are recorded by update lines. In this way, all the samplings are done in the restricted Hilbert space and it contains the cluster update scheme for constrained systems~\cite{ZY2020improved} and works at all temperatures. Proper finite size scaling analysis can then be carried out to explore phase transitions and critical phenomena.

We set the initial state as one of the two LN patters with the same probability to satisfy the two dimer per site constraint in our QMC simulation. The random initialization has no influence on the QMC results\cite{yanFully2022}. Our simulations are performed on the square lattice with periodic boundary condition and system sizes  $L=8,12,16,20,24,$ and $28$, while setting the inverse temperature $\beta=2L$ and using $10^5$ Monte Carlo samplings to obtain average values of the observables in all calculations.

\section{Phase diagram and transitions}
To explore the phase diagram of the Hamiltonian in Eq.~\eqref{eq:eq1}, we first use three order parameters given by
\begin{equation}
\begin{aligned}
\phi_{1} &= \frac{1}{N}|N^c_{\overdimer}-N^c_{\ohordimer}| \\
\phi_{2} &= \frac{1}{N}|N^c_{\verdimers}-N^c_{\hordimers}| \\
\phi_{3} &= \frac{1}{N}|N^c_{\sql\ or \sqr}-N^c_{\sqb\ or \sqt}|,
\end{aligned}
\end{equation}
where $N^c$ is the number of a specific dimer pattern (including \overdimer, \ohordimer, \verdimers\,  \hordimers, \sql, \sqr, \sqb, and \sqt) on all plaquettes of the dimer covering $c$. $\phi_{1}$ and $\phi_{2}$ are the single- and dimer-pair rotational symmetry breaking order parameters, respectively. As shown in Fig.~\ref{fig:fig1}(b), both of them can be used to detect the LN phase where the global lattice rotation symmetry is broken, and $\phi_{3}$ can be used to further detect whether in each plaquette there is a further local rotational symmetry breaking [as show in Fig.~\ref{fig:fig2} (b) and as explained below, $\phi_3$ is not the order parameter of the RP phase but nevertheless helps to find an irrelevant $\mathbb{Z}_4$ instability therein close to the RK point].

Our data in Figs.~\ref{fig:fig2}(a), ~\ref{fig:fig2}(b), and ~\ref{fig:fig2}(c) reveal $\phi_{1}$, $\phi_{2}$ and $\phi_3$ detect the LN-RP first order phase transition at $V=-0.35(3)$ [and we further show a coexistence of both phases in the histogram in Figs.~\ref{fig:fig4}(c) and ~\ref{fig:fig4}(d)]. And the small peak in $\phi_3$ around $V\sim 0.9$ in Fig.~\ref{fig:fig2}(c) seems to indicated the spontaneous breaking of an extra $\mathbb{Z}_2$ rotational symmetry with respect to the centers of a plaquette. However, this is a manifestation of an interesting RG flow path starting from the RK toward the RP fixed point, where along the way the Nambu-Goldstone (NG) fixed point of broken U(1) symmetry (which contains the further $\mathbb{Z}_2$ rotational symmetry inside a plaquette) is passed by [see the flow diagram in Fig.~\ref{fig:fig4}(a)].
But since this RG flow is triggered by a (dangerously) irrelevant operator $\cos(4\pi h)$---translating from the dimer configuration to height variable (see below)---close to the RK point, the finite size scaling of $\phi_{3}$ versus $1/L$, shown in the inset of Fig.~\ref{fig:fig2}(c), clearly demonstrates that its peak value around $V\sim 0.9$ eventually extrapolates to zero as a function of $L$ in the TDL.

\begin{figure}[t]
	\centering
	\includegraphics[width=1\columnwidth]{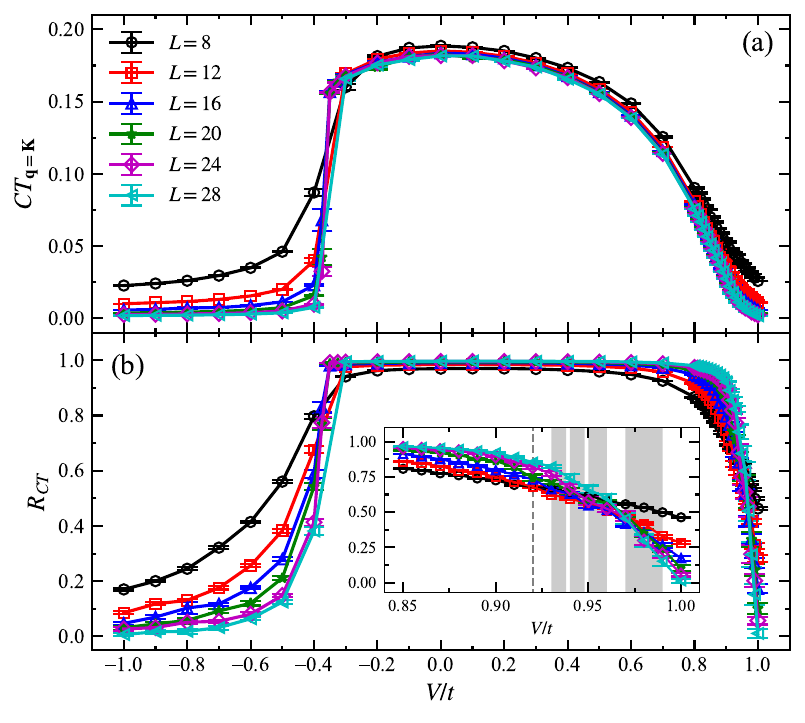}
	\caption{Off-diagonal correlation function and correlation ratio. (a) The static structure factor $CT_{\mathbf{q=K=(\pi,\pi)}}$. The finite value of $CT$ in the region of $-0.35(3)<V<1$ indicates the RP phase. (b) The correlation ratio $R_{CT}$. The inset shows the slow drifts in the crossing region of $R_{CT}$ between consecutive system sizes $L$, the crossing regions are denoted by the vertical gray bars,  and the drift moves to the RK point at $V=1$ in the TDL.}
	\label{fig:fig3}
\end{figure}

Now that the RP phase does not break the lattice rotation but the translational symmetry, we find the true order parameter of the RP phase shall be computed from the correlation function of the off-diagonal $t$ terms of the Hamiltonian, $CT$\,$\equiv$\,$\langle t_i t_j \rangle$ where $i$ is the position of the lower-left site of each plaquette. $CT$ captures the resonance of the parallel dimer pairs within a plaquette, hence the name RP phase. Since the RP order is invisible from the diagonal probes such as $\phi_{1,2,3}$ and can only be seen via off-diagonal $CT$, it is a new quantum state with {\it hidden} order which is of interest to Rydberg experiments. The translational symmetry breaking in RP presents itself as $\mathbf{K}=(\pi,\pi)$ in the static structure factor $CT_{\mathbf{q}=\mathbf{K}}$ shown in Fig.~\ref{fig:fig3} (a). Besides, its correlation ratio $R_{CT}=1-CT_{\mathbf{q'}}/CT_{\mathbf{q}=\mathbf{K}}$, where $CT_{\mathbf{q'}}$ is the average structure factor of the four momenta around $\mathbf{K}$ point, is shown in Fig.~\ref{fig:fig3}(b).

One sees in the LN phase that $R_{CT}$ extrapolates to zero in the TDL and it stays close to one in the RP phase, and the discontinuous jump signifies the first order transition at $V=-0.35(3)$. The interesting behavior is close to the RK point, where we observe the crossing of the $R_{CT}$ between consecutive $L$-s [denoted by the vertical dashed bars in the inset of Fig.~\ref{fig:fig3} (b)]. These crossing regions drift toward the RK point at $V=1$ as $L$ increases. This is another signature of the rich RG flow path close to the RK point with the dangerously irrelevant NG operator close by, and although these data demonstrate the RP phase within  $-0.35<V<1$, the relations between different fixed points originated from the RK point in the phase diagram of Fig.~\ref{fig:fig1}(b) clearly deserve further explanations from a field theoretical perspective, as we now turn to.

\begin{figure}[htp!]
	\centering
	\includegraphics[width=1\columnwidth]{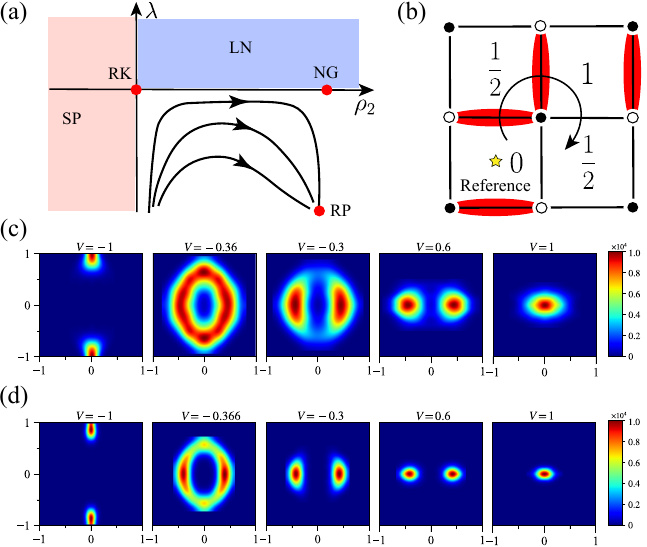}
	\caption{(a) Schematic RG flow. When $\rho_2>0$ the flow is toward the RP fixed point with $\lambda<0$, passing the irrelevant Nambu-Goldstone (NG) fixed point, which signifies a further $\mathbb{Z}_2$ rotational symmetry breaking inside each plaquette, as denoted by the small peak close to $V\sim 0.9$ in Fig.~\ref{fig:fig2}(c). The LN ($\lambda>0$) and SP ($\rho_2<0$) regions are also denoted. (b) Height representation. The construction of height pattern at a $2\times 2$ unit of plaquettes. Histograms of height variable order parameter for (c) $L=8$ and (d) $L=16$ system. The histogram in the LN phase, RP phase, and at the RK point are shown. At $V=-0.36$ in (c) and $V=-0.366$ in (d), the coexistence of the LN and RP phases at the first order phase transition manifests. As for $V=-0.366$ in $L=16$ system, the histogram demonstrates the first-order transition between the LN and RP phases more clearly, which shows it is not a finite-size effect.}
	\label{fig:fig4}
\end{figure}

\section{Effective action and renormalization-group analysis}
It is known that the QLM has an effective height field action in the continuum limit~\cite{henleyRelaxation1997}. The rule for constructing the height covering~\cite{davidCoulomb2003,neilSynchronization2019,neilInteracting2020} is illustrated in Fig.~\ref{fig:fig4}(b). We first choose one plaquette as the reference with zero height, then taking a clockwise (counterclockwise) movement around sites with ``magnetic charge" $+2 (-2)$, let the height increase by $1/2$ when crossing an occupied dimer bond and decrease by $1/2$ for an empty bond. In this way, dimer coverings $\{c\}$ can be translated to height coverings $\{h\}$, and we further define $h_I$ as the average of the four plaquettes centering the ``magnetic charge" site.

\begin{figure*}[htp!]
	\centering
	\includegraphics[width=0.9\textwidth]{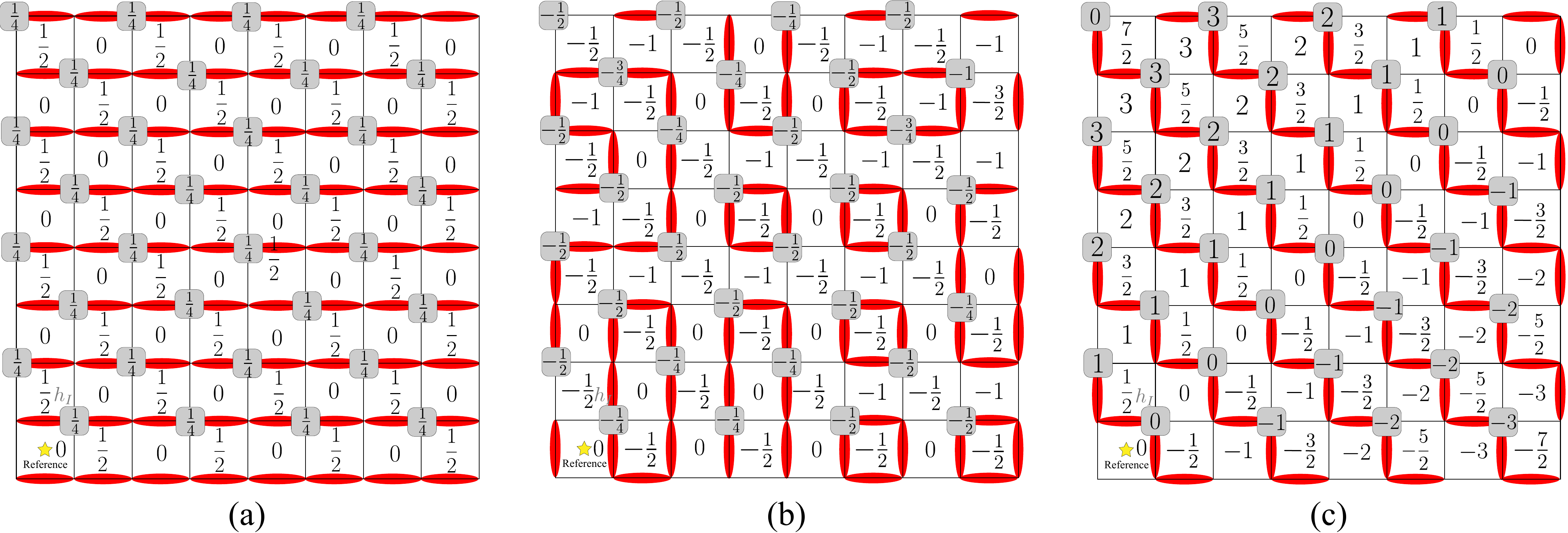}
	\caption{Height pattern for dimer coverings when (a) $V=-1$, (b) $V=0.3$, and (c) $V>1$ on a $8\times 8$ lattice, corresponding to the LN, RP, and SP states discussed in the main text. The height value in each plaquette is given by the rule shown in  Fig.~\ref{fig:fig4}(b), which shows the clockwise movement around the $+2$ ``magnetic charge". The values in the gray rectangle are the average height $h_I$ of the $2\times 2$ unit. The values of the height variable in the LN, RP, and SP states are consistent with the RG analysis. In the SP, the height variable is periodic with respect to $L$.}
	\label{fig:sfig1}
\end{figure*}

Here we show the height patterns for $V=-1$ (LN  phase), $V=0.3$ (RP phase), and $V>1$ (SP) in Fig.~\ref{fig:sfig1}. For the LN and RP phases, we choose one dimer configuration in our QMC simulations for each phase, the distribution of the average height variable $h_I$ consist with the histogram indicated in Fig.~\ref{fig:fig4} (c), where $h_I=\pm 1/4$ in the LN phase and $h_I=\pm 1/2$ or $0$ in the RP phase. For the SP, we give an ideal dimer configuration to show the height pattern with the Lagrangian parameter $\rho_2<0$.  Since the $\nabla h$ is a finite value and $h$ is periodic with respect to $L$. We shall construct the height pattern in a finite lattice with caution. When the reference height is fixed with zero in the first plaquette, we first give the height variables of the diagonal plaquettes, then determine its value of the upper (lower) triangular plaquettes according to the upper left (lower right) height variable of the diagonal plaquettes. In this way, we can find that along the diagonal of our lattice $|\nabla h|=1$ and the period of the height variable is $L/2$.

{Near the RK point, the QLM can be mapped to a quantum version of the famous six-vertex model  \cite{ardonneTopological2004}, which can be described by the following Lifshitz type of Lagrangian~\cite{fisherStatistical1961,kasteleynThe1961,temperley1961dimer,fisherStatisticalII1963,di1987relations,ardonneTopological2004,fradkinBipartite2004,aletClassical2006}}
\begin{align}\label{effectiveaction}
\mathcal{L}=\frac{1}{2}(\partial_{\tau}h)^2+\frac{1}{2}\rho_2 (\nabla h)^2+\frac{\kappa^2}{2} (\nabla ^2 h)^2+\lambda \cos(4\pi h),
\end{align}
 where $\rho_2\propto -(V-V_c)$ with $V_c=1$ the RK point. The RG flow diagram of the Lagrangian is shown in Fig.~\ref{fig:fig4}(a). When $\rho_2>0$, the spatial derivative of the height field $h$ is suppressed. This favors a spatial homogenous $h_I$, which is indeed the case in the RP phase [see Fig.~\ref{fig:sfig1}(b)].
 When $\rho_2<0$, $\nabla h$ jumps immediately to its cutoff value, which corresponds to the SP phase in Fig.~\ref{fig:sfig1}(c).
 The height variable is periodic with the identification $h=h+1$. The $\lambda \cos(4\pi h)$ term therefore breaks the symmetry from $U(1)$ to $\mathbb{Z}_2$, with $\lambda<0$ and the $\lambda>0$ favors the RP and LN phases, respectively.
 The RK point, located at $\rho_2=\lambda=0$, is a quantum critical point with dynamical critical exponents $z=2$, at which the action is invariant under the scaling symmetry $\tau \rightarrow l^2 \tau$ and $\vec{x} \rightarrow l \vec{x}$ and the operator $ \cos(4\pi h)$ is irrelevant, indicating an emergent $U(1)$ symmetry~\cite{kivelsonTopology1987,ardonneTopological2004}.

We calculate the histogram of the height variable $\langle \frac{4}{N}\sum_{I}\cos(2\pi h_I), \frac{4}{N}\sum_{I}\sin(2\pi h_I) \rangle$, i.e., average over the lattice , for a $L=8$ system, which are shown in Fig.~\ref{fig:fig4} (c). In the LN phase, the order parameter is $(0,\pm 1)$ with $h_I=\pm 1/4$, while that is $(\pm 1,0)$ in the RP phase with $h_I=\pm 1/2$ or $0$. At $V=-0.36$, the coexistence of the LN and RP phases at the first order phase transition manifests. From the histogram, it is clear that as we approach the RK point, the $\mathbb{Z}_2$ symmetry gets enhanced to $U(1)$, although it is still clear there is a $\mathbb{Z}_2$ anisotropy at $V=1$ in the finite size histogram in Fig.~\ref{fig:fig4}(c) and ~\ref{fig:fig4}(d), consistent with our date in Figs.~\ref{fig:fig2}(c) and ~\ref{fig:fig3}(b).  
 
\begin{figure}[htp!]
	\centering
	\includegraphics[width=0.9\columnwidth]{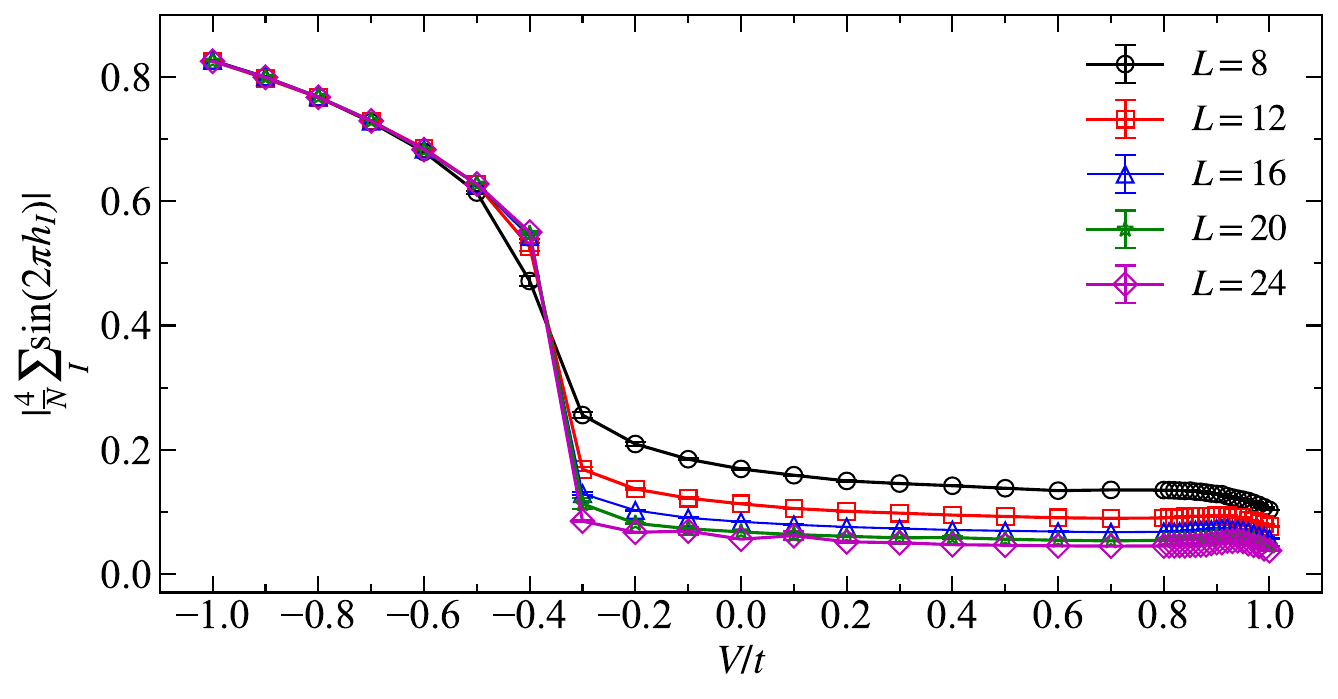}
	\caption{The height order parameter $|\frac{4}{N}\sum_i\sin(2\pi h_I)|$ for different system sizes, which shows the same behavior as the order parameters in Fig.~\ref{fig:fig2}. The mild peaks in the region of $0.9<V<1$ are the finite size effect which result in the anisotropy of the histogram shown in Fig.~\ref{fig:fig4}(c).}
	\label{fig:sfig2}
\end{figure}

The height order parameter $|\frac{4}{N}\sum_{I}\sin(2\pi h_I)|$ for different system sizes is shown in Fig.~\ref{fig:sfig2}, which indicates the same small peak in the region of $0.9<V<1$ as the order parameters in Fig.~\ref{fig:fig2}(c). $\frac{4}{N}\sum_{I}\sin(2\pi h_I)$ is the $y$ component of the histograms illustrated in Fig.~\ref{fig:fig4}(c), which corresponds to $\pm 1$ in the LN phase and $0$ in the RP phase. The mild peaks of the order parameter exhibit the dangerously irrelevant $\mathbb{Z}_2$ instability at $V\sim 0.9$ close to the RK point and will eventually vanish at the the thermodynamic limit. Our RG flow of Eq.~\eqref{effectiveaction}, shown schematically in Fig.~\ref{fig:fig4}(a) and date of $|\frac{4}{N}\sum_{I}\sin(2\pi h_I)|$ in Fig.~\ref{fig:sfig2}, show the anisotropy is a finite size effect.

 The Lagrangian Eq.~\eqref{effectiveaction} with positive $\rho_2$ and $\lambda=0$ is precisely the Nambu-Goldstone (NG) Lagrangian that describes the spontaneous symmetry breaking phase of $U(1)$ symmetry \cite{nambuQuasi1960,goldstone1961field,goldstoneBroken1962}. When $\lambda\neq0$, the NG fixed point is unstable under RG flow.  At a fixed coupling, the QLM with increasing lattice size can be viewed as an RG flow. Near the RK point, the $\lambda \cos(4\pi h)$ term is irrelevant, so that the RG flow lingers around the $\lambda$ close to zero region (in particular, near the NG fixed point) for a very long RG time. This means that a finite-sized QLM will likely be described by Eq.~\eqref{effectiveaction} with small $|\lambda|$. Notice the energy difference between the minimum and the maximum of the $\lambda \cos(4\pi h)$ is proportional to $|\lambda|$. When $|\lambda|$ is small (and $\lambda<0$), quantum fluctuations can easily drive vacuum from the RP phase with $h=\pm1/2$ or $0$ to other values. This explains the non-vanishing $\phi_3$ at $V\sim  0.9$ in Fig.~\ref{fig:fig2}(c), the drift of the crossing point in Fig.~\ref{fig:fig3}(b), and the anisotropy of histogram in Fig.~\ref{fig:fig4}(d) at $V=1$ in finite size data. Similarly,  the expectation value of $|\frac{4}{N}\sum_{I}\sin(2\pi h_I)|$, which is shown in Fig.~\ref{fig:sfig2}, shows a mild peak in the region of $0.9<V<1$. A similar RG flow with crossover behavior was observed in a three dimensional classical clock model~\cite{shaoMonte2020} or in the deconfined quantum critical point where a dangerously irrelevant second length scale is found to mask the still unsettled fixed point~\cite{senthilQuantum2004,shaoQuantum2016,Nahum2015a,zhaoScaling2022,wangScaling2022,liuFermion2022}.

The above mentioned RG structure relies on the fact of the operator $\cos(4\pi h)$ being irrelevant at the RK point. The scaling dimension of $\cos(4\pi h)$ is related to the ``Luttinger parameter'' $\kappa$. Different with the QDM case ($\kappa=2 \pi $) which can be mapped to the free fermion~\cite{kasteleynThe1961,temperley1961dimer}, QLM can be mapped to the famous six-vertex model~\cite{gier2009fully}. There are in total six possible dimer configurations at each site. It has been known that the six-vertex model can be described as free compact boson models in two dimensions~\cite{di1987relations}: 
\begin{equation}
\begin{aligned}\label{vac}
\langle vac| &\mathcal{O}[h(x_1)] \ldots \mathcal{O}[h(x_n)] |vac \rangle = \\
&\frac{1}{\mathcal{Z}}\int [\mathcal D h]\mathcal{O}[h(x_1)] \ldots \mathcal{O}[h(x_n)] e^{-\kappa \int dx^2 (\nabla h)^2},
\end{aligned}
\end{equation}
where $\kappa=2\pi r^2$, with $r$ being the radius of the compact boson in the convention of Ref.~\cite{GINSPARG1988153}.  All correlation functions on the RK vacuum is equivalent to the six-vertex model with all six types of vertices appearing at equal probability (which are usually called the ice point, referring to the ice rule of water ice by Pauling~\cite{paulingThe1935}). At the ice point, one gets $\kappa=\pi/3$~\cite{di1987relations}. The 2D quantum version of this type of model was derived in Ref.~\cite{ardonneTopological2004} and leads to Eq.~\eqref{effectiveaction}.
The conformal field theory analysis of the free compact boson tells us that the scaling dimension of $\cos(2\pi n h)$ [and $\sin(2\pi n h)$] is $\Delta_n=\frac{ n^2 \pi}{2 \kappa}=\frac{3n^2}{2}$ \cite{Ginsparg:1988ui}. Clearly $\Delta[{\cos(4\pi h)}]=6$, which is irrelevant.

 \begin{figure}[htp!]
	\centering
	\includegraphics[width=1\columnwidth]{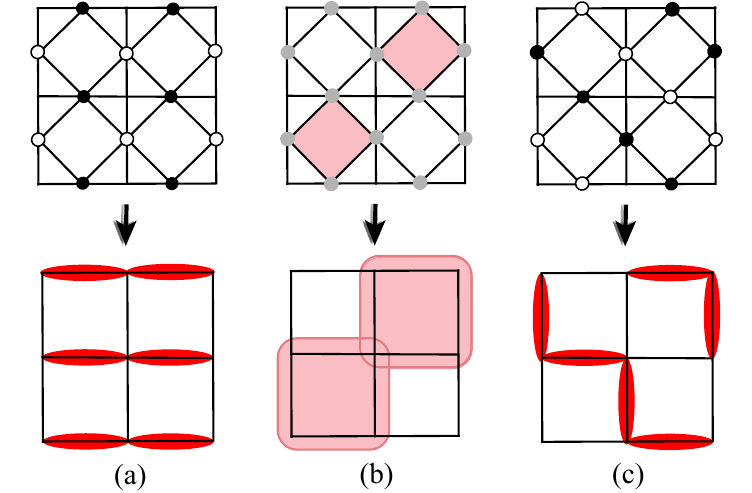}
	\caption{Mapping of checkerboard lattice Rydberg atom array to QLM on square lattice. The correspondences between the Rydberg atom configuration and the (a) LN phase, (b) the RP phase, and (c) the SP of QLM model are demonstrated. The solid (hollow) circle represents the presence (absence) of a Rydberg atom on the bonds of the checkerboard lattice, and the grey circle in (b) denotes the resonating of the Rydberg atoms in each tetrahedron.}
	\label{fig:fig5}
\end{figure}

\section{Experimental proposal and Discussion}
The QLM on square lattice can be realized in the Rydberg arrays experiments, which have recently been utilized to probe topological order in QDMs~\cite{satzingerRealizing2021,semeghiniProbing2021}. The effective Hamiltonian of Rydberg arrays is~\cite{saffmanQuantum2010, browaeysMany2020}. $H_{\rm R}=h \sum_{i} \sigma^x_i -\mu \sum_{i} n_i + V\sum_{i>j}\frac{n_i n_j}{|i-j|^6}$, where $i$ and $j$ are the site labels, $n_i=0,~1$ is the density operator to probe the ground state or Rydberg state, respectively, and $\sigma^x$ is the tunneling term to connect the two states. If we only consider the nearest-neighbour (NN) interactions, and the NN runs over the bonds of the 2D checkerboard lattice~\cite{shannonCyclic2004} in Fig.~\ref{fig:fig5}. Under the competing between the Rydberg blockade (favoring one particle in Rydberg radius) and chemical potential (inducing more occupied particles), the Rydberg atom configurations will obey the ``ice rule" in certain parameter region---every state with exactly two occupied and two empty sites per tetrahedron (cross linked plaquette)~\cite{paulingThe1935}, i.e., there are two particles in the Rydberg radius. The kind of local constraint has been realized both in experiment~\cite{semeghiniProbing2021} and numerical simulations of Rydberg Hamiltonian~\cite{samajdarQuantum2021,yanEmergent2023}.

In the $h \ll V$ case, any excitation breaks the ``ice rule" will lead a huge energy cost at the scale of $V$. Therefore, the quantum fluctuations within the ``ice rule" becomes the low energy excitations, that is the 4th order term of the $h$, i.e., $h^4 (\sigma^+_i\sigma^-_j\sigma^+_k\sigma^-_l+\sigma^-_i\sigma^+_j\sigma^-_k\sigma^+_l)$, where the $i,~j,~k,~l$ are the sites labeled in a plaquette. Thus, the low energy effective model in this case is the QLM in Eq.~\eqref{eq:eq1}. In this way, the LN, RP and SP states and their interesting phase transitions, the {\it hidden} nature of the RP as a new quantum state of matter, and the intricate RG flow with a marginal operator in the effective height action, can all be investigated in the Rydberg atom experiments.

{\noindent\it Acknowledgements---}We thank Fabien Alet for valuable discussions on the phase diagrams of QDM and QLM over the years. X.X.R, Z.Y, and Z.Y.M acknowledge support from the Research Grants Council of Hong Kong SAR of China (Grants No.~17303019, No.~17301420, No.~17301721, No.~AoE/P-701/20, No.~17309822), the ANR/RGC Joint Research Scheme sponsored by Research Grants Council of Hong Kong SAR of China, and French National Research Agency (Project No. A\_HKU703/22), the K. C. Wong Education Foundation (Grant No. GJTD-2020-01), and the Seed Funding ``Quantum-Inspired explainable-AI'' at the HKU-TCL Joint Research Centre for Artificial Intelligence. Y.Q. acknowledges support from the the National Natural Science Foundation of China (Grants No.~11874115 and No.~12174068). J.R. is supported by Huawei Young Talents Program at IHES. Y.C.W. acknowledges support from Zhejiang Provincial Natural Science Foundation of China (Grant No. LZ23A040003) and Beihang Hangzhou Innovation Institute Yuhang. We thank HPC2021 system under the Information Technology Services and the Blackbody HPC system at the Department of Physics, the University of Hong Kong for their technical support, and generous allocation of CPU time.

\bibliographystyle{apsrev4-1}
\bibliography{main}

\end{document}